%% file: main.tex
\documentclass[fleqn,usenatbib]{mnras}
\input{Content/preamble.tex}
\input{Content/header.tex}

\begin{document}
\label{firstpage}
\pagerange{\pageref{firstpage}--\pageref{lastpage}}
\maketitle

\input{Content/1-Abstract.tex}

\input{Content/2-Introduction.tex}
\input{Content/3-Method.tex}

\input{Content/4-Extension.tex}
\input{Content/5-Simulations}

\input{Content/6-KV450data}

\input{Content/7-Results}
\input{Content/8-Conclusion.tex}
\input{Content/acknowledgement.tex}
\input{Content/data_availability.tex}

\input{Content/9-Appendices}

\bibliographystyle{mnras}
\bibliography{ads}


\bsp	
\label{lastpage}
\end{document}

%% file: Content/preamble.tex
\usepackage[T1]{fontenc}
\usepackage{ae,aecompl}

\usepackage{array}
\usepackage{graphicx}	
\usepackage[nice]{nicefrac}
\usepackage{bm}
\usepackage{textcomp}
\usepackage{hyperref}
\usepackage{natbib}
\usepackage{amsmath}
\usepackage{graphicx}
\usepackage[colorinlistoftodos]{todonotes}
\usepackage{xcolor}
\usepackage{multirow}
\usepackage{todonotes}
\usepackage{float}
\usepackage{booktabs}
\usepackage{subcaption}
\usepackage{lineno}
\usepackage{seqsplit}
\usepackage{verbatim}
\usepackage{ulem}
\usepackage{tikz}
\usepackage{calrsfs}
\DeclareMathAlphabet{\pazocal}{OMS}{zplm}{m}{n}
\usetikzlibrary{positioning}
\usetikzlibrary{shapes,snakes}
\usepackage{amsfonts}

\usepackage{color}
\definecolor{green}{rgb}{0.0,0.5,0.0}
\definecolor{darkblue}{rgb}{0.,0.,0.4}
\definecolor{darkred}{rgb}{0.5,0.,0.}
\hypersetup{urlcolor=darkred, citecolor=blue, linkcolor=blue}

\usepackage{mathptmx}

\definecolor{ForestGreen}{RGB}{34, 139, 34}

\newcommand{\bs}[1]{\boldsymbol{#1}}
\newcommand{\mc}[1]{\pazocal{#1}}
\newcommand{\tm}[1]{\textrm{#1}}

%% file: Content/header.tex
\title[KV450 cosmology with BHM $n(z)$]{KiDS+VIKING-450 cosmology with Bayesian hierarchical model redshift distributions}

\author[G. T. Kyriacou et al]{George T. Kyriacou,$^{1}$
Arrykrishna Mootoovaloo,$^{1,2}$
Alan F. Heavens,$^{1}$\thanks{E-mail: a.heavens@imperial.ac.uk}
Andrew H. Jaffe,$^{1}$
\newauthor
Florent Leclercq,$^{3}$
Konrad Kuijken$^{4}$
\\
$^{1}$Imperial Centre for Inference and Cosmology (ICIC), Imperial Astrophysics, Imperial College, Blackett Laboratory, Prince Consort Road, London, SW7 2AZ, UK\\
$^{2}$ Oxford Astrophysics, Department of Physics, University of Oxford, Denys Wilkinson Building, Keble Road, Oxford, OX1 3RH, UK\\
$^{3}$ CNRS \& Sorbonne Université, UMR 7095, Institut d'Astrophysique de Paris, 98 bis boulevard Arago, F-75014 Paris, France\\
$^{4}$ Leiden Observatory, Leiden University, P.O. Box 9513, 2300 RA Leiden, The Netherlands}

\date{Accepted XXX. Received YYY; in original form ZZZ}

\pubyear{2026}

%% file: Content/1-Abstract.tex
\begin{abstract}
Tomographic redshift distributions from photometric data are crucial ingredients in cosmic shear analysis, since they are required for the theoretical calculation of the signal based on the redshift distribution of the galaxies where the shear field is sampled.
In this paper, we develop as a proof of concept Leistedt et al.'s template-based Bayesian Hierarchical Model framework into an application to weak lensing data by sampling the redshift distributions of the galaxies in the KiDS+VIKING-450 survey.  We also use a principal component analysis to provide a set of representative templates drawn from a large superset.  For computational tractability, subsets of $10^5$ galaxies are chosen to determine the redshift distributions, and we test the sensitivity of the cosmological inference to the subset chosen, finding it to be subdominant compared to the statistical error.  We marginalise over the inferred redshift distributions and find that the Bayesian method increases the clustering parameter compared with previous studies, alleviating the $S_8$ tension with Planck, where $S_{8}\equiv\sigma_{8}\sqrt{\Omega_{\tm{m}}/0.3}=0.756\pm 0.039$, assuming flat $\Lambda$CDM. 
The tension with Planck for this survey is reduced from $2.3\sigma$ to $1.9\sigma$. We also infer a value for the matter density, $\Omega_{\tm{m}}=0.31\pm 0.10$. 
\end{abstract}

\begin{keywords}
gravitational lensing: weak -- cosmology: cosmological parameters -- techniques: photometric -- methods: data analysis, statistical
\end{keywords}

%% file: Content/2-Introduction.tex
\section{Introduction}

Testing cosmological models through the inference of cosmological parameters is one of the main goals of large galaxy surveys. One observable is cosmic shear, the distortion of images of galaxies due to weak gravitational lensing along the line-of-sight by cosmic structure \citep[for reviews see e.g.][]{2001PhR...340..291B,2008PhR...462...67M,2015RPPh...78h6901K}. Due to the large volumes required to obtain a measurable signal, this is usually accomplished with broad-band photometric surveys rather than spectroscopy, leading to some challenges in determining the redshift distribution of the sources, which is required to compute the 
cosmic shear or convergence power spectrum $P_\kappa$. This is sensitive particularly to the median redshift $z_m$ \citep{2006MNRAS.366..101H, 1997ApJ...484..560J, 1997A&A...322....1B}, but in detail the full $n(z)$ is required.  Since cosmic shear increases with both source redshift and power spectrum amplitude, any bias in the median redshift results in an opposite bias in the inferred power spectrum amplitude. The demand for accurate $n(z)$ is exacerbated through the use of tomographic bins, which are needed to increase the information content.  Methods have been developed to estimate photometric redshifts from low spectral resolution data, such as template fitting \citep{2011Ap&SS.331....1W} and more specifically Bayesian Photometric Redshifts (BPZ) \citep{2000ApJ...536..571B}. Template fitting is the matching of photometric data to that of redshifted spectra that have been multiplied by the appropriate filter response and a free amplitude, and integrated. 

As accurate as these approaches may be in giving estimates and redshift posteriors for {\it individual} galaxies, they do not translate to developing usable and accurate $n(z)$ for the galaxies in the survey, which is what is used for cosmology inference. Previous surveys such as CFHTLenS and DES used the stacking of individual redshift posterior distributions \citep{2012MNRAS.421.2355H, 2016PhRvD..94d2005B}. However, this approach magnifies the scatter of the underlying $n(z)$ and may not be the most suitable way to infer the distribution \citep{2005MNRAS.359..237P, 2019PASJ...71...43H}. This may lead to biases at a level intolerable for current and future surveys \citep{2015APh....63...81N, 2016MNRAS.463.3737C, 2022ApJ...928..127M, 2023arXiv230204507E}. Some methods do not give uncertainties in the values of $n(z)$, which are then often treated by introducing wholesale shifts in the tomographic redshift distributions, which cannot reflect the full variability of the redshift distributions, so marginalisation over the redshift distribution uncertainty is only approximate.

Current surveys have attempted to alleviate this problem through the use of a complementary spectroscopic survey. One such method is the weighted direct calibration, referred to as DIR \citep{2020A&A...633A..69H,2021A&A...650A.148S,2023MNRAS.522.5037R}. First proposed by \cite{2008MNRAS.390..118L}, the method takes accurate spectroscopic redshifts and calibrates the final $n(z)$ by re-weighting the spectroscopic redshift distribution so that the weighted spectroscopic survey matches the magnitude and colour distributions of the photometric survey. This method has been shown to be both accurate and gives uncertainties, through the use of bootstrap samples of said surveys, in their estimations. Unfortunately, this accuracy relies on a spectroscopic survey that covers the full range of magnitudes, redshifts and types of galaxies in the photometric sample, a luxury one will not have as photometric surveys become deeper in redshift and broader in scope \citep{2009arXiv0912.0201L, 2019A&A...627A..59E} as noted by \cite{2020A&A...638L...1J}. Furthermore, uncertainties are produced but samples cannot be taken in a strictly Bayesian sense. 

Cosmic shear analysis from the first year data of the Hyper Suprime-Cam  \citep[HSC;][]{2019PASJ...71...43H} used a similar approach to DIR for obtaining redshift distributions, the main difference being that overlapping spectroscopic surveys \citep{2015MNRAS.452.2087L, 2013A&A...559A..14L} do not have galaxy numbers large enough for the reweighting process to represent accurately properties of the photometric sources. As an alternative, the HSC survey uses the COSMOS 30-band photo-z catalogue \citep{2009ApJ...690.1236I} - a significantly larger number of bands than KV450, at the expense of an extra source of variability in the errors in the COSMOS photo-zs.  

An alternative technique for inferring the $n(z)$ redshift distributions is the cross-correlation (CC) method developed by \cite{2008ApJ...684...88N}, with which we will also make comparisons. 
This method uses the cross-correlation functions between photometric and spectroscopic objects to finding the photometric redshift distributions.  

An increasingly widespread technique is self-organising maps (SOM) \citep{2019MNRAS.489..820B,2020A&A...637A.100W}, where colours are mapped to cells, and spectroscopic data used to assign redshifts to cells.  Recently, \cite{2023MNRAS.524.5109R} developed a method to draw posterior samples of redshift distributions for the HSC Y3 analysis. In the DES Y3 analysis, the SOM method was used to generate samples of the redshift distributions and these were then used to obtain cosmological constraints from galaxy clustering and weak lensing \citep{2021MNRAS.505.4249M, 2022PhRvD.105b3520A}. A similar approach using SOM was used in the KiDS-1000 analysis \citep{2021A&A...645A.104A}. A challenge of methods that rely on spectroscopic data is covariate shift - mismatched spectroscopic and photometric samples, for which care needs to be taken, such as in the weighting scheme of StratLearn \citep{Autenrieth}.  This challenge motivates a method for obtaining Bayesian $n(z)$ distributions and uncertainties from photometric data alone.

In this paper, we adapt
the method presented by \citet{Leistedt2016}, who presented a Bayesian Hierarchical Method (BHM) with the main goal of inferring the redshift distribution of the population from which they are drawn \citep[see also][]{2023ApJS..264...29A, 2023ApJS..264...23L}.
For weak lensing, current surveys do not account for source clustering, so we instead infer the redshifts of the background galaxies where the shear field is sampled. While individual galaxy samples of template type, magnitude and redshift are generated as an intermediate step in the Gibbs scheme of \cite{Leistedt2016}, they are not retained or used as a direct inferential output in that framework. The innovation here is to retain these samples and use them to construct tomographic redshift histograms for each observed galaxy subsample, shifting the inferential target from the population distribution to the distribution of the specific galaxies used in the shear analysis. Hence in this work we develop a practical algorithm to sample the  
individual redshifts of the galaxies 
in the survey, rather than the population from which they are drawn. 
We also refine the choice of templates by developing a PCA-based method for grouping similar templates. These are empirical, largely based on observed galaxy spectra, in contrast to modelling the spectra with stellar population models, see e.g. \cite{2023ApJS..264...23L,popcosmos}.  

In \S\ref{sec:method}, we develop the method and test on simulated data in \S\ref{sec:simulations}.   The KV450 data are described in \S\ref{sec:kids_vikings_450_analysis}, and the redshift distributions derived.   In \S\ref{sec:results}, we present the cosmological results and discuss various challenges, and finally we present our conclusions in \S\ref{sec:conclusions}.  Appendices present the simulation details and the (standard) background theory for completeness.

%% file: Content/3-Method.tex
\section{The Bayesian Hierarchical Model}
\label{sec:method}

The Bayesian approach used here is similar to that presented by \cite{Leistedt2016}.  We parametrize the galaxies in terms of a template $t$, redshift $z$, and ($r$-band) apparent magnitude $m$, with the ultimate goal being to sample the population distribution $p(t,z,m)$, finally marginalising over the templates and magnitudes to produce samples of the redshift distributions for each tomographic bin.  To do this we alternately sample (a) the individual galaxy $t,z,m$ values,  and (b), the parameters describing the population distribution of $t,z,m$, all conditioned on the measured galaxy fluxes $\{\hat F_b\}$.  

\begin{figure}
\centering
\includegraphics[scale=0.8]{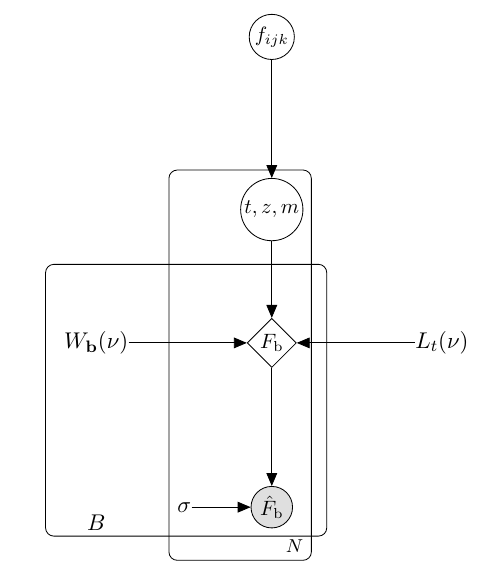}
\caption{Hierarchical forward model.  $f_{ijk}$ is the fraction of galaxies in bins of template $t$, redshift $z$ and magnitude $m$, labelled by $i,j,k$ respectively. $t,z,m$ values are drawn for all $N$ galaxies, from the population specified by $f_{ijk}$, and noise-free fluxes $F_b$ computed  for the $B=9$ photometric bands $b=1\ldots B$.  The bands are defined by transmissions $W_b(\nu)$, and the templates by $L_t(\nu)$.  The observed fluxes $\hat F_b$ include flux errors given by $\sigma$. Empty circles indicate quantities drawn from probability distributions, the diamond shows a deterministic relation, the filled circle represents (fixed) observed data, and quantities on their own are considered known precisely. The plates labelled by $B$ and $N$ indicate multiple instances. This can be applied to the whole sample or a subsample assigned (e.g. by BPZ) to a tomographic bin.}
\label{fig:BHM}
\end{figure}

For each galaxy, the measured flux $\hat F_b$ in a photometric band labelled by $b$ is assumed to have a gaussian error distribution, \begin{equation}
    \hat F_b \sim {\mc N}(a F_b,\sigma^2_b),
    \label{eq:fluxerror}
\end{equation}
where $\sigma_b^2$ is the flux variance of the galaxy in band $b$, and $a$ is an amplitude. $F_b$ is the noise-free flux, given by integrating the redshifted rest-frame template spectrum $L_t(\nu)$ multiplied by the filter response $W_b(\nu)$ over frequency, 
\begin{equation}
    F_b(t,z) = \frac{1+z}{4\pi D_{L}^2(z)}\int_{0}^\infty \frac{d\nu}{\nu}L_t(\nu)[\nu(1+z)]W_b(\nu),
    \label{eq:meanflux}
\end{equation}
where $D_L(z)$ is the luminosity distance.  Note that the templates have an arbitrary normalisation, so an amplitude $a$ is introduced, and set by dividing the unnormalised flux (Equation \ref{eq:meanflux}) by the unnormalised $r$-band flux, and then multiplying $F_b$ by the desired $r$-band flux, determined by $m$.

We discretise the 3D distribution $p(t,z,m)$ in a 3D histogram of template, redshift and magnitude, labelled by indices $i,j,k$, with $f_{ijk}$ representing the fraction of the population residing in each bin.   Note that $i$ is a genuinely discrete label, whereas $j$ and $k$ label continuous parameters which are binned (note that $j$ represents a much finer binning than that of the tomographic redshift distribution of interest; for reference, the binning in $m$ is from $m=19$ to $m=25$ in steps of 0.1, and in $z$, steps of 0.05 up to $z=3$). The survey population 3D histogram then approximates the joint distribution $p(t,z,m)$, integrated over the bins of redshift and magnitude:
 \begin{equation}
f_{ijk} = \int^{z_{j,\rm{max}}}_{z_{j,\rm{min}}} \int^{m_{k,\rm{max}}}_{m_{k,\rm{min}}}p(t_i,z,m)\,dz\, dm.
\end{equation}
Similarly, the conditional sampling distribution of the fluxes of galaxies in a bin is
\begin{equation}
    p( \hat{F}_b|ijk)=\int^{z_{j,\rm{max}}}_{z_{j,\rm{min}}} \int^{m_{k,\rm{max}}}_{m_{k,\rm{min}}}p(\hat{F}_b |t_i,z,m)\,dz\,dm,
\end{equation}
where the probability distribution for $\hat F_b$ in the integrand is given by Equation \ref{eq:fluxerror}. The probability distribution for the set of measured fluxes of a galaxy, $\{ \hat F_b\}$, is a product of the individual band likelihoods, assuming independence.

Each coefficient $f_{ijk}$ is a population parameter to be inferred and the sampling captures any inter-bin correlations, which may be important for parameter inference. 
We assume a uniform prior for $f_{ijk}$, subject to constraints $0\le f_{ijk}\le 1$ and $\sum_{ijk}f_{ijk} = 1$ (the Dirichlet distribution).  The uniform prior in template index $i$ is a prior on the galaxy types in the template set used, which we discuss in \S\ref{ssec:template_choice}. The relevant directed acyclic graph is shown in Figure \ref{fig:BHM}.

We employ a 2-step Gibbs sampler, alternating between sampling the set of individual galaxy ($g$) templates, redshifts and magnitudes, $\{(t, z, m)_g \}$, and the population parameters $\{ f_{ijk} \}$.  The sampling differs slightly from the efficient implementation of \cite{Leistedt2016}, which describes a Gibbs sampling implementation that marginalizes over the individual galaxy properties $(t,z,m)$, and samples the underlying population $\{f_{ijk}\}$ from which the galaxies were drawn. However, for use in tomographic weak lensing analysis, we require the properties not of the population from which the observed galaxies are drawn, but the redshifts of the galaxies that are actually in the sample, since these are the points where the shear field is sampled.  Sampling both the coefficients $\{f_{ijk}\}$ and the individual properties $(t,z,m)_g$ for all galaxies allows us to use the latter samples, with each galaxy labelled by its tomographic bin, to construct the histograms for each tomographic bin separately. 

We note that the redshift distributions inferred by this method differ conceptually from the population-level distributions that appear in the standard derivations of the lensing power spectra \citep[e.g.][]{2001PhR...340..291B}, which are fixed properties of the survey selection function and binning, whereas our method samples the histograms of redshifts of specific galaxies in the survey.  These two quantities differ in principle since our method includes the effect of source clustering on the average redshift distribution .  However, we argue that for the purposes of a wide-field survey such as KV450, this distinction is negligible in practice for two reasons.  Firstly, these deviations are small, as the volumes in which we sample the redshift distributions are large, averaged over the several hundred square degrees of the survey, in bins that are of order 50 Mpc thick, and this is a completely subdominant effect.  Secondly, source clustering is a small effect even for much higher precision surveys such as Euclid \citep{Linke2025}.  The distinction between the two inferential targets is therefore not of practical significance for the present analysis, though we acknowledge it as an interesting conceptual question and an important direction for future work as survey volumes and statistical precision increase.

\subsection{Tomographic bin sampling}

For each sample (which is a full histogram and set of properties for all observed galaxies), we finally allocate each galaxy to its predetermined (by BPZ) tomographic bin and counting these gives a sample of the numbers, $n_{ijk,{\rm bin}}$, from which we marginalise over template $i$ and magnitude $k$ to create a histogram of the redshifts. This produces samples of $n(z)_{\rm bin}$ which can be marginalised over in the cosmological analysis; we also have access to the posterior distribution of the properties of each source which may separately be of interest.

Note that there are three independent uses of ``binning'' in our model. The initially BPZ tomographic binning procedure is deterministic, and performed outside of and essentially separately from this hierarchical model. In practice, BPZ may use some of the same templates as in our set $\{t_i\}$, but it is important to understand that BPZ is used {\it only} as a mechanism to divide the sample into tomographic bins. Inevitable errors in BPZ photometric redshifts mean that the actual distribution of redshifts will extend beyond the  nominal redshift limits of each bin, but this does not matter.  The crucial requirement is that we have sufficient samples to represent the actual redshift distributions. 

To set up the Gibbs sampling, where we alternately sample from (a) the $i,j,k$ for each galaxy and (b) the population parameters $f_{ijk}$, we start by using Bayes theorem: assuming independent galaxy observations, the joint distribution (see \cite{Leistedt2016} for full details) is:
\begin{equation}
    p(n_{ijk},\{f_{ijk}\}|\{ \hat{F}_b \}_g) \propto p(\{f_{ijk}\})\prod^{N_{\rm{gal}}}_{g=1}p(\{ \hat{F}_b \}_g|ijk) p(ijk|\{f_{ijk}\}).
\end{equation}
For given $f_{ijk}$, the Gibbs sampling step then samples $i,j,k$ for each galaxy, with probability
\begin{equation}
    p(ijk|\{f_{ijk}\},\{ \hat{F}_b \}_g) = f_{ijk} \times p(\{ \hat{F}_b \}_g|ijk).
    \label{eq:Gibbs}
\end{equation}
Note that no tomographic binning is done at this stage. Counting galaxies in each $ijk$ then gives a sample $n_{ijk}$ for the whole population of sources.  Using this $n_{ijk}$ one draws samples for $f_{ijk}$ following a multinominal distribution:

\begin{equation}
\begin{split}
   p(\{f_{ijk}\}|\{n_{ijk}\}) = &
   (N_{\rm{gal}}+N_tN_zN_m-1)!\delta_D(1-\sum_{ijk}f_{ijk}))\\
   &\prod^{N_t}_{i=1}\prod^{N_z}_{j=1}\prod^{N_m}_{k=1}
   \frac{\Theta(f_{ijk})f_{ijk}^{n_{ijk}}}{n_{ijk}!}.
   \label{eq:Gibbs2}
\end{split}
\end{equation}

So far, the approach described above follows \cite{Leistedt2016}, whose main goal was to sample the underlying distribution $f_{ijk}$ from which survey galaxies are drawn, given the selection effects. However, for a weak lensing analysis, we require the redshift distribution 
of the individual galaxies that occupy each tomographic bin of that survey, not the population from which they are drawn, which requires a different approach.  Each galaxy has a label determined by BPZ, so the algorithm is simply to add up the number of galaxies in each bin $ijk$ for each tomographic bin label for each realization.  We then marginalise over template $i$  and magnitude  $k$ to obtain a sample of each tomographic redshift distribution.

%% file: Content/4-Extension.tex
\subsection{Template set}
\label{ssec:template_choice}

 One of the key assumptions of this model is that all galaxies within the survey are represented closely enough by one of the templates in the set used in the analysis. 
 Thus, one of the issues is that of template incompleteness; one requires a large enough set of templates that they cover the spectral range of the photometric data. This would suggest using a very large template set.  Our prior is that each template is equally probable, so the template set determines the prior.  For example, if a set contained six elliptical templates for each spiral, the prior would favour elliptical galaxies. This imbalance can propagate into the template posterior, if there is not enough data to override the prior.

\begin{figure}
\noindent \begin{centering}
\includegraphics[width=0.45\textwidth]{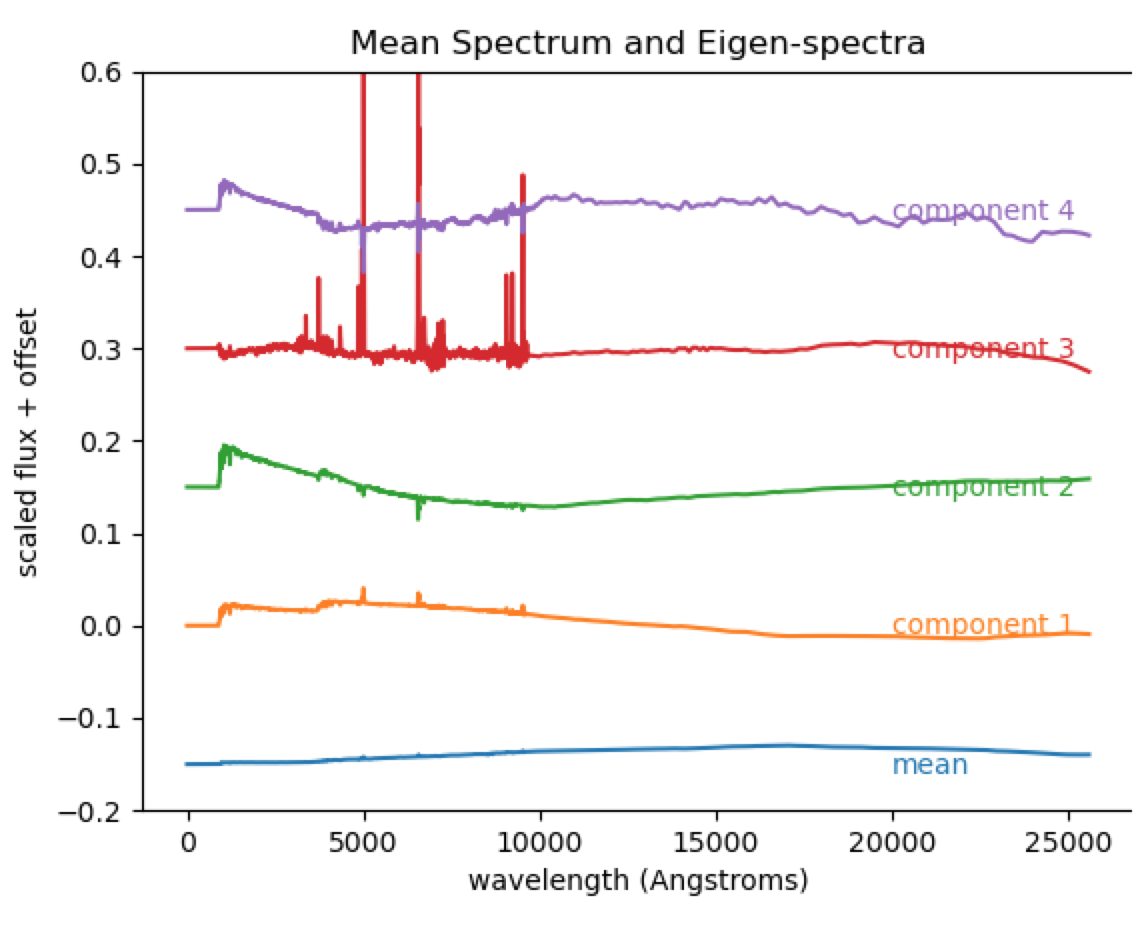}
\includegraphics[width=0.4\textwidth]{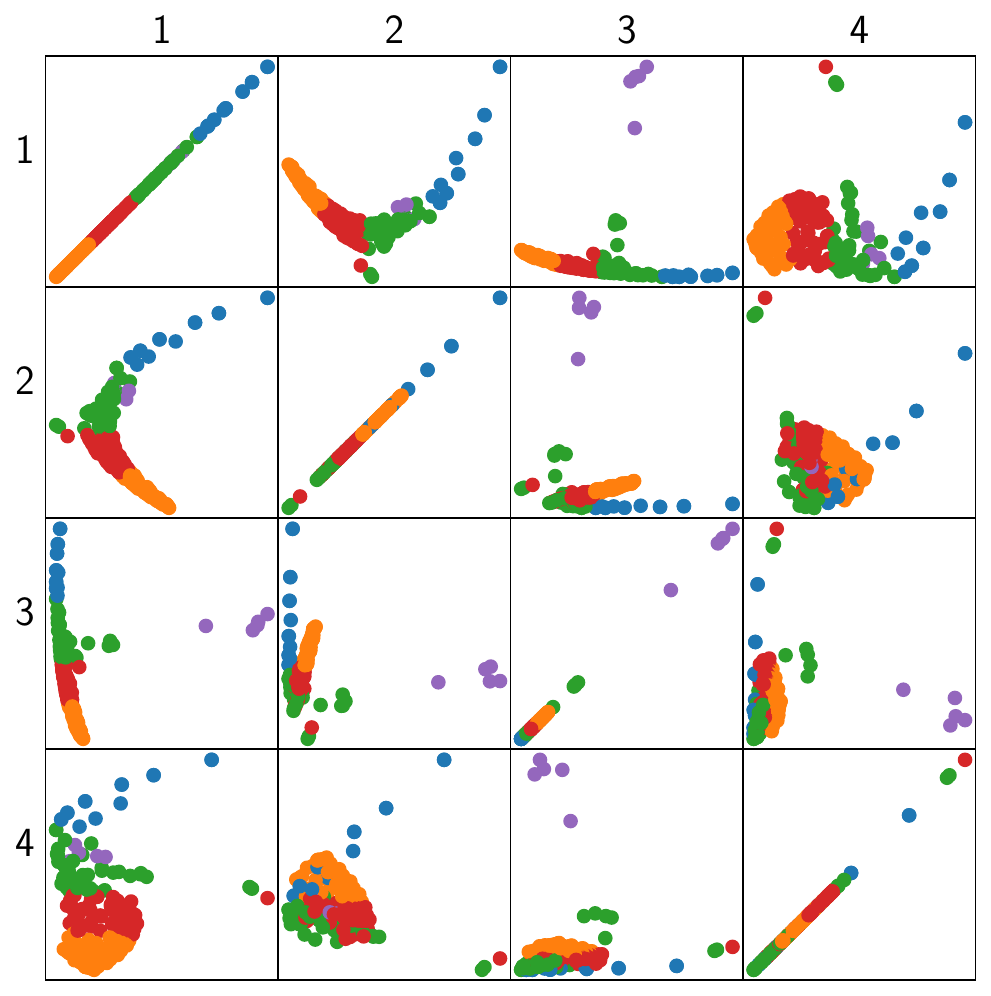}
\par\end{centering}
\caption{\textit{Top}: Scatter plot of examples of the first 4 principal components extracted from a large template set. The flux has been scaled and offset for visualisation purposes. \textit{Bottom}: a 4x4 scatter plot of all templates plotted against component number. The colours indicate the 5 different sets where the $k$-means clustering algorithm placed the template.} 
\label{fig:PCA}
\end{figure}

We have constructed a template superset of around 300 galaxies from the BPZ source at \href{https://www.stsci.edu/~dcoe/BPZ/}{\tt{https://www.stsci.edu/$\sim$dcoe/BPZ/}}. However, using so many templates makes the code slow. Therefore, we used smaller sets of templates which still capture the colours of the survey galaxies. In addition to choosing random sets of templates, we also created representative sets by using principal component analysis (PCA; using {\tt{sklearn}}) and $k$-means clustering in the 4D space of the largest 4 principal components, to divide the templates into 5 groups.  We do not use templates constructed from the principal components themselves, but use PCA only to divide the templates into similar groups, from each of which we choose 10 templates at random, thus avoiding over-representation of similar templates which may be over-abundant in the complete template sample.   This turned out to be an effective way to reduce the number of templates required to obtain stable results. 
The four components are shown in the upper panel of Fig. \ref{fig:PCA} in which obvious characteristics of certain templates, such as emission lines, can be seen.  In the lower panel, we see the groupings in different colours.  This device allows us to sample representative templates without biasing the results towards types that are overrepresented in the rather heterogeneous complete template set.
The method can be adapted to other template choices, such as those generated from theoretical models, as in the pop-cosmos Bayesian Hierarchical model programme \citep{2023ApJS..264...29A,2024ApJS..274...12A,2024ApJ...975..145T,2025arXiv250612122T,2025arXiv250920430D}, or a mixture of the two.

%% file: Content/5-Simulations.tex
\section{Tests on Simulations}
\label{sec:simulations}

To test the validity of the method and the new implementation, sets of simulated redshift distributions were created. We draw $N_{\rm{gal}}= 2 \times 10^6$ $r$-band magnitudes from the $r$-band distribution of the KV450 survey. Types were drawn uniformly and independently of magnitude from the full set of over 300 templates, so there is a level of model misspecification in that this is a wider set than the PCA template set used for analysis.  The redshifts of the simulated sources were drawn from the following conditional distribution:
\begin{equation}
    p(z|m_k) = \frac{4}{5}\mc{N}(\mu_{1,k},\sigma^{2}_{1,k}) + \frac{1}{5}\mc{N}(\mu_{2,k},\sigma^{2}_{2,k})
\end{equation}
\noindent and $\mu,\sigma$ are the means and variances of the Gaussian distributions.  Lastly $5\%$ of the simulated population were given redshifts independently of magnitude and type following a uniform distribution, to test the method's ability to assign correctly sources that are outliers from the main population.  We emphasise that this is not meant to be an approximation to the real galaxy distribution, but is a test case of a non-trivial distribution that is not unimodal.

The redshift distribution parameters are both chosen to increase with magnitude bin ($m_k$) in order to replicate some of the behaviour seen in the BPZ priors. Explicitly, we take
\begin{eqnarray}
\mu_{1,k} &=& 0.20 + 0.25 \, m_k \\
\mu_{2,k} &=& 0.8 + 0.25  \, m_k \\
\sigma^2_{1,k} &=& 0.05 + 0.05\, m_k \\
\sigma^2_{2,k} &=& 0.18 +  0.02\, m_k .
\end{eqnarray}
when $m_k=[19,20,21,22,23,24]$. 

\begin{figure*}
\noindent \begin{centering}
\includegraphics[width=0.85\textwidth]{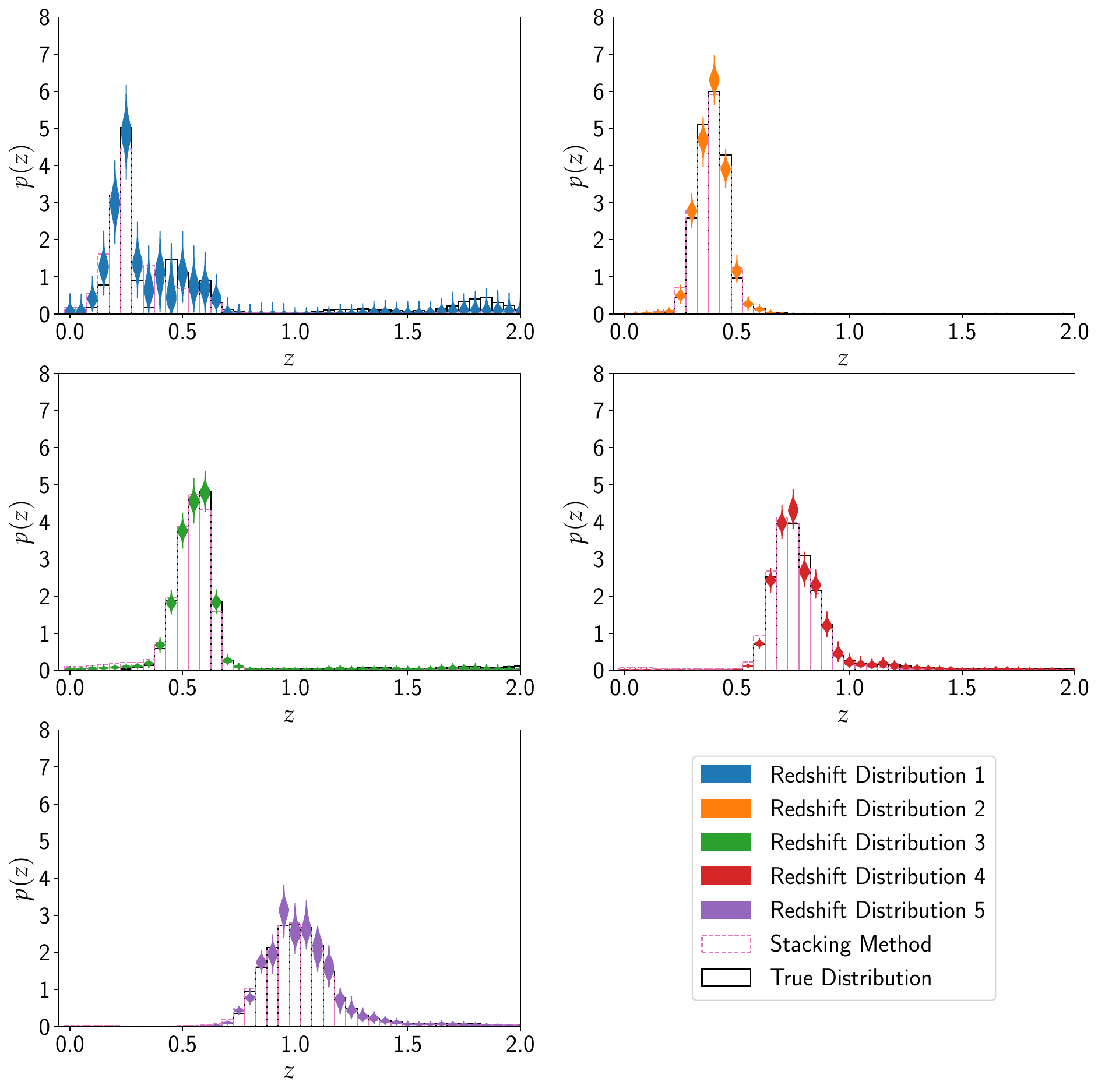}
\par\end{centering}
\caption{\label{fig:sim_stack_true} 
The tomographic redshift distributions generated using the simulated data. The redshift intervals for each tomographic distribution are: $0.1<z\leq 0.3$, $0.3 < z \leq 0.5$, $0.5 < z \leq 0.7$, $0.7 < z \leq 0.9$ and $0.9<z\leq 1.2$. The shaded region on each bin (violin plot) indicates the uncertainty associated with the inferred heights of the tomographic redshift distribution. The solid black bars indicate the true distribution while the coloured dashed bars represent the stacked likelihood.}
\end{figure*}
\begin{table}

\begin{tabular}{|c|c|c|c|c|c|}
    \hline
    &\multicolumn{2}{c|}{BHM}& \multicolumn{2}{c|}{Stacked} \\
    \hline
     $z_B$ range & $\Delta z_m$  & $\Delta \sigma^2_{z}$ & $\Delta z_m$  & $\Delta \sigma^2_{z}$ \\
    \hline
    Full Survey & 0.004  & -0.003 &  -0.011 & 0.035\\
    $0.1 < z_B \leq 0.3$ & 0.003  & 0.015 & 0.004  & 0.213\\
    $0.3 < z_B \leq 0.5$ & -0.002 &  -0.0005 & 0.004 & 0.003 \\
    $0.5 < z_B \leq 0.7$ & 0.004 & -0.003 & 0.014 & 0.076 \\
    $0.7 < z_B \leq 0.9$ & 0.001 & -0.002 & 0.012 & 0.035\\
    $0.9 < z_B \leq 1.2$ & 0.005 & -0.003 & 0.004 & 0.035  \\
    \hline
\end{tabular}
\caption{The average error in median redshift and variance from the simulated data, using the two analysis methods, namely the stacked likelihood (SL) and the BHM model of this paper. Note that the BHM performs better in almost every measure.}
\label{tab:z_mean}     
\end{table}

To simulate magnitudes and their errors and galaxy weights, we have used distributions from the KV450 dataset.  Details of the procedure are given in Appendix A.

To test the method, we compare the inferred redshift distributions with the true simulated distributions.  3000 samples were collected on the $t,z,m$ cell allocations of each source. Weights were then applied to each source before summing to obtain the redshift distributions. Fig. \ref{fig:sim_stack_true} compares the stacked likelihood with the BHM, showing good recovery of the input redshift distributions.  Visually there is not too much difference, but a quantitative analysis shows that the BHM is a significant improvement over the stacked likelihood, which is expected given that it should provide a principled statistical solution.

We focus on the median redshifts  $z_m$, since they are mostly responsible for the inferred value of $\sigma_8$. We show the differences between the inferred medians and the truth in Table  \ref{tab:z_mean}, shows that the BHM provides a more accurate value than the stacked likelihood.  We also see that the variance is also recovered more accurately.

In more detail, the average $\chi^2$ across the 5 tomographic bins is calculated to be $\overline{{\chi}^{2}} = 102$ and $\overline{{\chi}^{2}} = 18$ for the stacked likelihood and BHM model respectively. Thus the BHM model again shows significant improvement. It should also be noted that both approaches perform worse in the $5^{\tm{th}}$ bin (with the stacked likelihood and BHM having $\chi^2_{5^{\tm{th}}} = 308$ and $\chi^2_{5^{\tm{th}}} = 34$ respectively). A possible consequence of this is that the photometric data are less informative at higher redshifts where more filters would be of use.

%% file: Content/6-KV450data.tex
\section{KiDS + VIKING 450 analysis}
\label{sec:kids_vikings_450_analysis}
\subsection{Data}

The data used for the cosmological parameter inference and demonstration of the application of this Bayesian hierarchical model were the KiDS+VIKING-450 data set\footnote{\url{https://kids.strw.leidenuniv.nl/DR3/kv450data.php}} \citep{Wright2018,Kannawadi2018}. This is a combination of optical (KiDS) and near infrared (VIKING) photometric data. KiDS is a wide angle photometric survey which use the OmegaCAM camera within the VLT Survey Telescope \citep{DeJong2012}. With a target area of $\sim$ 1500 deg$^2$, the survey intends to reveal cosmological information through the large scale distribution of matter. Although multiple data releases have taken place \citep{DeJong2015,DeJong2017} KiDS-450 refers to DR3 of which 450 deg$^2$ of sources in the $ugri$ bands have become available. 

Infrared data is collected from the VISTA Kilo-degree infrared Galaxy survey which used the Visible and Infrared CAMera (VIRCAM) on ESO’s 4m VISTA telescope \citep{Venemans2015}. The planned target is for VIKING to cover $\sim$ 1350 deg$^2$ of sky across the $ZYJHK_s$ IFR bands. Due to incompleteness in the coverage of VIKING, in order to ensure that the combined data set of KV450 is fully covered in all nine bands, it consequently has a reduced unmasked area of 341 deg$^2$. 

The specific data used was the Gaussian Aperture and PSF magnitudes (GAaP), a post processing stage of data reduction unchanged from previous releases \citep{Kuijken2008,Kuijken2015}. Although other forms of multiband data are available the GAaP magnitude/fluxes are the default used by KiDS. As the details of the photometric data are more relevant to the Bayesian Hierarchical Model than for DIR, a further data filtering was undertaken, following \cite{Bilicki2017}, to which we refer the reader for the rationale and discussion. Sources that do not include magnitude errors in every band are removed as well as those requiring GAaP\_Flag\_ugriZYJHKs=0. This requirement reduced the KV450 area by $\sim 5 \%$. Unlike the analysis used by KiDS for both their DIR and BPZ for full redshift distribution and tomographic bin allocation, we have not used the GAaP magnitudes, rather the GAaP fluxes and errors have been used. These are available in the KiDS data set as `FLUX\_GAaP\_b' and `FLUXERR\_GAaP\_b', where `b' labels the 9 filters. The fluxes also need to be corrected before analysis using offset and extinction correction terms:

\begin{equation}
    F_{\tm{corrected},b} = F_b \times 10^{\frac{2}{5}(\text{extinction}_b-\text{offset}_b)},
    \label{eq:correction}
\end{equation}
\noindent where $F_{\tm{corrected},b}$ is the final flux and $F_b$ is the flux in the catalogue.  A similar modification is made to the flux errors. The extinction term is taken directly from `EXTINCTION\_b' column of the data set. The magnitude offset is a per-tile correction. Sources are grouped by the tile of sky in which they were observed, this is found under `THELI\_NAME'. Sources within the same tile have the same offset which is calculated as the median value of the following quantity:
\begin{equation}
    \text{offset}_{b} = \text{MAG\_GAaP\_b} + 2.5\log_{10}(\text{FLUX\_GAaP\_b}).
\end{equation}
\noindent The median is used since there is typically a small number of undetected or unobserved sources, which don't have individual offsets.  The tomographic bin allocation is determined by the BPZ values reported in the original data `Z\_b' and the weights are taken directly from `recal\_weight'.  50 PCA templates were used for the analysis as described in \S\ref{ssec:template_choice}.

\subsection{Template and galaxy subsample sensitivity}

\begin{figure*}
    \centering
    \includegraphics[width=0.9\textwidth]{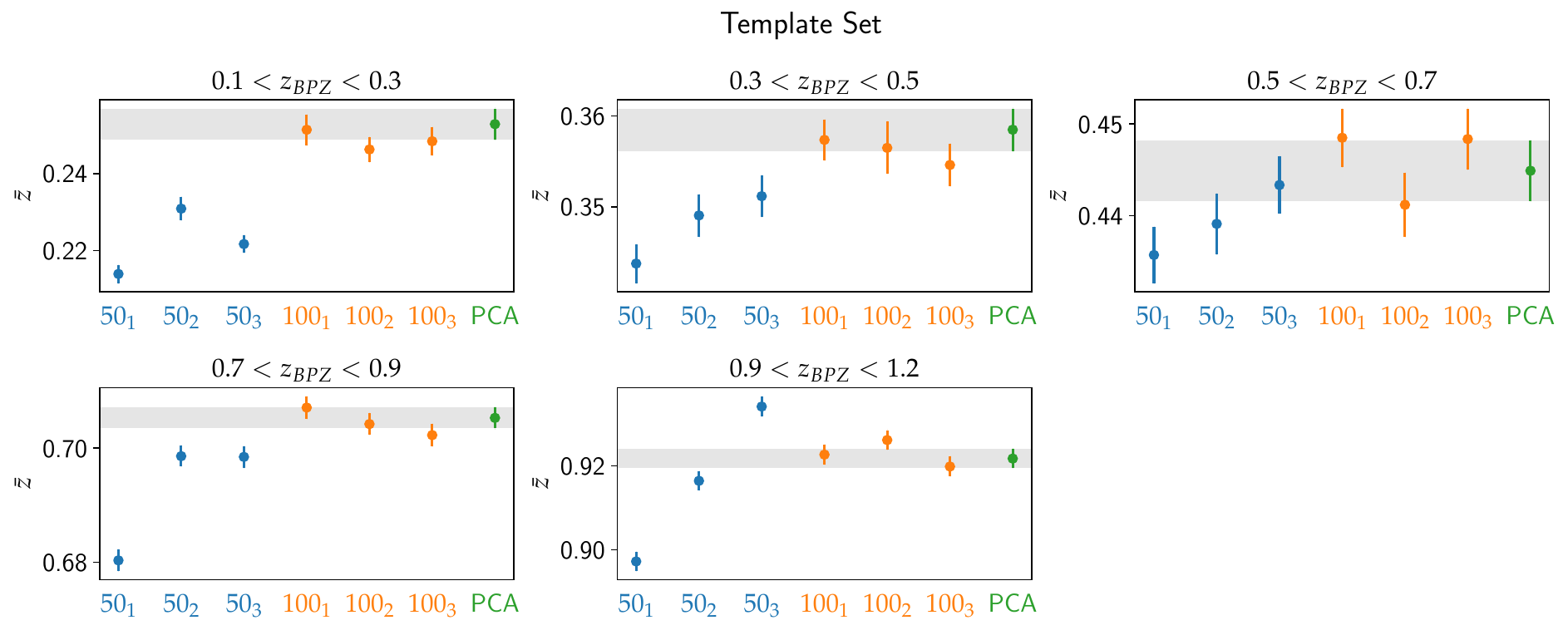}
    \caption{The median tomographic redshifts based on the redshift distributions of the KV450 data and their standard deviations, using 6 random template sets (3 containing 50 templates, denoted by $50_{n}$ and 3 containing 100 templates, denoted by $100_{n}$). These sets are compared to the same result from the 50 templates selected in roughly equal numbers from each of the clusters of templates grouped by clustering in the first four principal components space (`PCA', in green).  Note that 50 random templates gives widely varying median redshifts, whereas with 100 templates, the median redshifts are much more stable.  However, the 50 more controlled representative PCA-selected templates yield a median redshift that agrees well with the 100-template random sets, so we can use this smaller, faster PCA set for KV450 analysis.}
    \label{fig:mean_template}
\end{figure*}

To examine the issue of template sensitivity, 3 groups of randomly chosen 50 template sets and 3 groups of randomly chosen 100 template sets were used to produce redshift distributions for the 5 tomographic bins. The median redshift of the distributions was then compared to the median derived from 50 templates drawn using the PCA and $k$-means clustering technique presented in \ref{ssec:template_choice} to check for consistency. The median redshift was chosen as the comparison statistic due to its influence on the weak lensing power spectrum, since the main sensitivity of cosmological results to the redshift distributions is to the median values. For example, from Figure \ref{fig:mean_template}, we can see that when choosing 50 random templates, there is significant variability in the median redshift. However, the issue is no longer present when one chooses a set of 100, showing 100 templates to be an adequate number to achieve template insensitivity. However, when compared to the 50 templates present in the PCA analysis, one can see that the PCA 50-template set gives median redshifts that are in good agreement with the 100 template analyses, so for computational efficiency we use the PCA representative 50 template set for the cosmology analysis below.   

\begin{figure*}
\noindent \begin{centering}
 \includegraphics[width=0.8\textwidth]{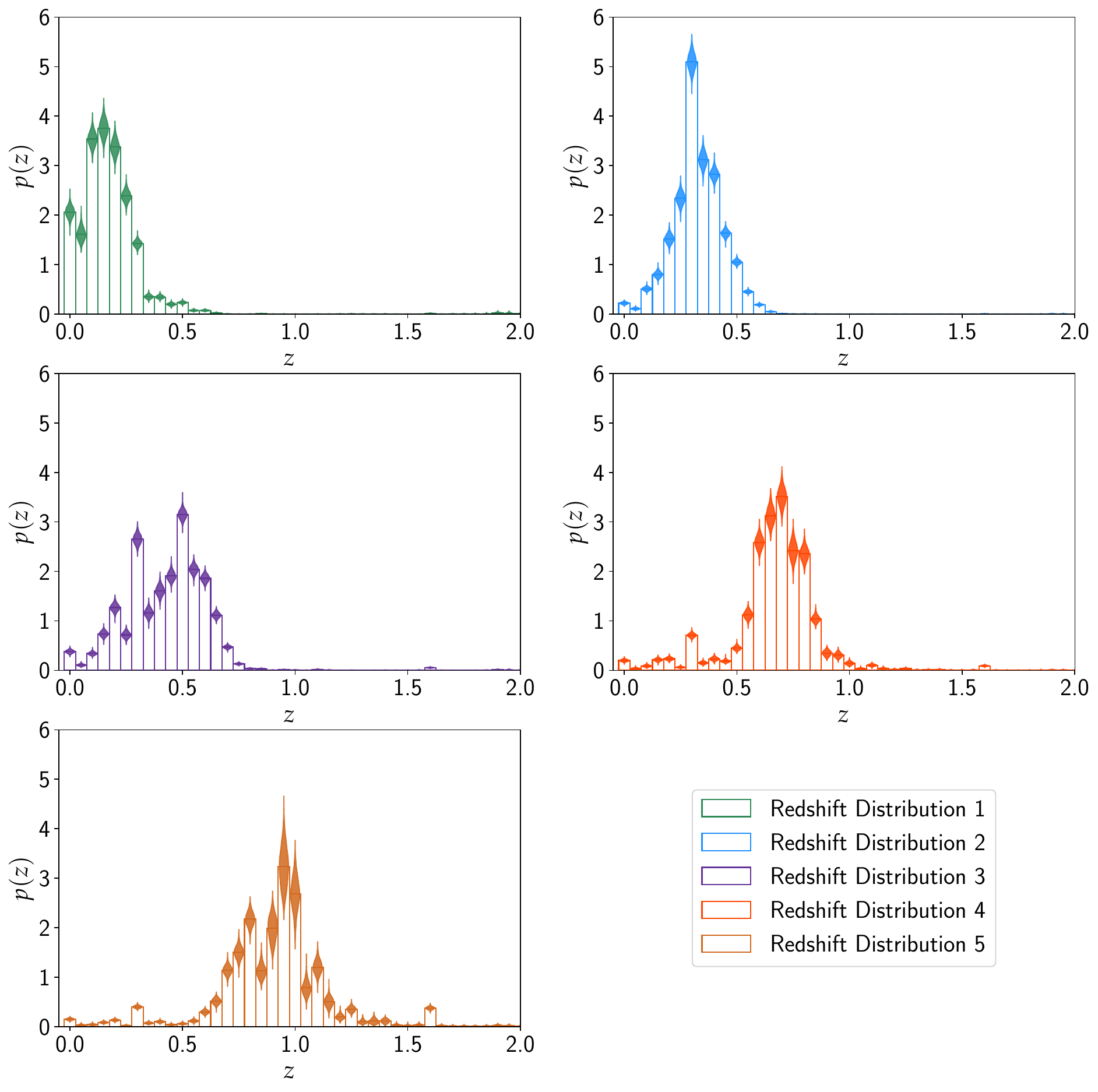}
\par\end{centering}
\caption{\label{fig:bhm_nz_1}The KV450 tomographic redshift distributions generated using the first set of galaxy samples. The nominal (BPZ) redshift intervals for each tomographic distribution are: $0.1<z\leq 0.3$, $0.3 < z \leq 0.5$, $0.5 < z \leq 0.7$, $0.7 < z \leq 0.9$ and $0.9<z\leq 1.2$. The shaded region on each bin (violin plot) indicates the uncertainty associated with the inferred heights of the tomographic redshift distributions. These are then propagated in the likelihood analysis when sampling the full posterior distribution of the cosmological and nuisance parameters.}   
\end{figure*}

Applying this method to all galaxies in the KV450 sample would be too slow, so we use a subset of galaxies to infer the redshift distributions.  To test sensitivity to the subset used, we randomly pick $10^{5}$ galaxies and infer their tomographic redshift distributions and repeat the experiment four times. We also generate a fifth set by making a ``combined set'', concatenating all the four subsamples.  

Taking the second half of the thinned MCMC chains generates 5000 and 20000 $n(z)$ samples for the individual and combined sets, which are then used to sample the cosmological and nuisance model parameters' posterior. We find only minor variations in the values of the parameters inferred, at the level of 0.01 in $S_8$ (see Fig. \ref{fig:1D_S8}), so in \S\ref{sec:results} we report only the results obtained with the combined 20000 $n(z)$ samples. 

Finally, in Appendix C we show the colours of the template set along with the colours of a random subset of the KV450 galaxies.  Given the large and variable errors on the photometry, it is difficult to make conclusive statements, but an expanded template set with wider colour coverage, particularly in the infrared, would be required for high-precision application of this method in future analyses.

\subsection{KV450 redshift distributions}

In Fig. \ref{fig:bhm_nz_1} we show a histogram of the combined samples of the tomographic redshift distributions for KV450, with shaded regions indicating the marginal posteriors of the histogram heights.  

\begin{figure*}
\noindent \begin{centering}
\includegraphics[width=0.8\textwidth]{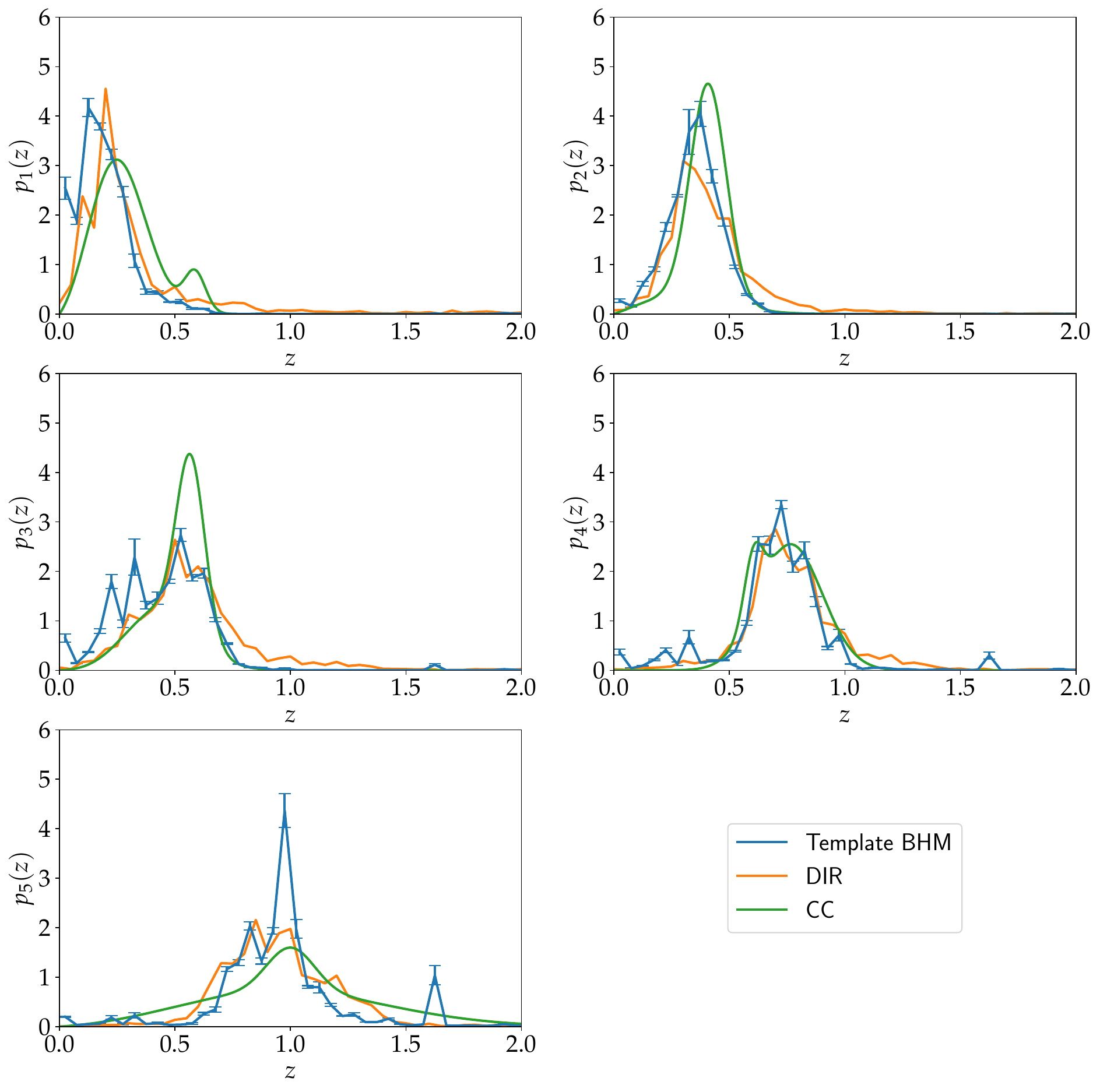}
\caption{\label{fig:bhm_nz_HH}The KV450 BHM tomographic redshift distributions (blue) compared with the DIR and CC methods of \citet{2020A&A...633A..69H}.  The blue points show the variability in the mean $n(z)$, determined from the different galaxy subsamples. The others are shown without errors.} 
\par\end{centering}
\end{figure*}

A comparison with other methods can be made by inspection of Appendix C of \cite{2020A&A...633A..69H}. We show a comparison with DIR and CC (after fitting with a Gaussian Mixture Model) in Fig. \ref{fig:bhm_nz_HH}. Since the methods are different, we do not expect them to agree in detail, but a comparison with the cross-correlation technique is useful to make. If there is a positive cross-correlation of the spectroscopic sample with the photometric sample at a given redshift, it is difficult to see how that could arise without some photometric galaxies being at those redshifts, so we might be concerned if the inferred BHM $n(z)$ was zero there.  Inspection of the CC curves of \cite{2020A&A...633A..69H} shows no evidence for non-zero CC redshifts where the BHM distributions vanish, except in the highest-redshift bin, where magnification could plausibly introduce correlations.  

The main differences are the less prominent high-redshift tail in the Bayesian method, giving rise to the higher $\sigma_8$ inferred here.

%% file: Content/7-Results.tex
\section{Cosmology results}
\label{sec:results}

\begin{figure*}
    \centering
    \subfloat[$\Omega_{\tm{m}}-S_{8}$ plane]{{\includegraphics[height=0.30\textheight]{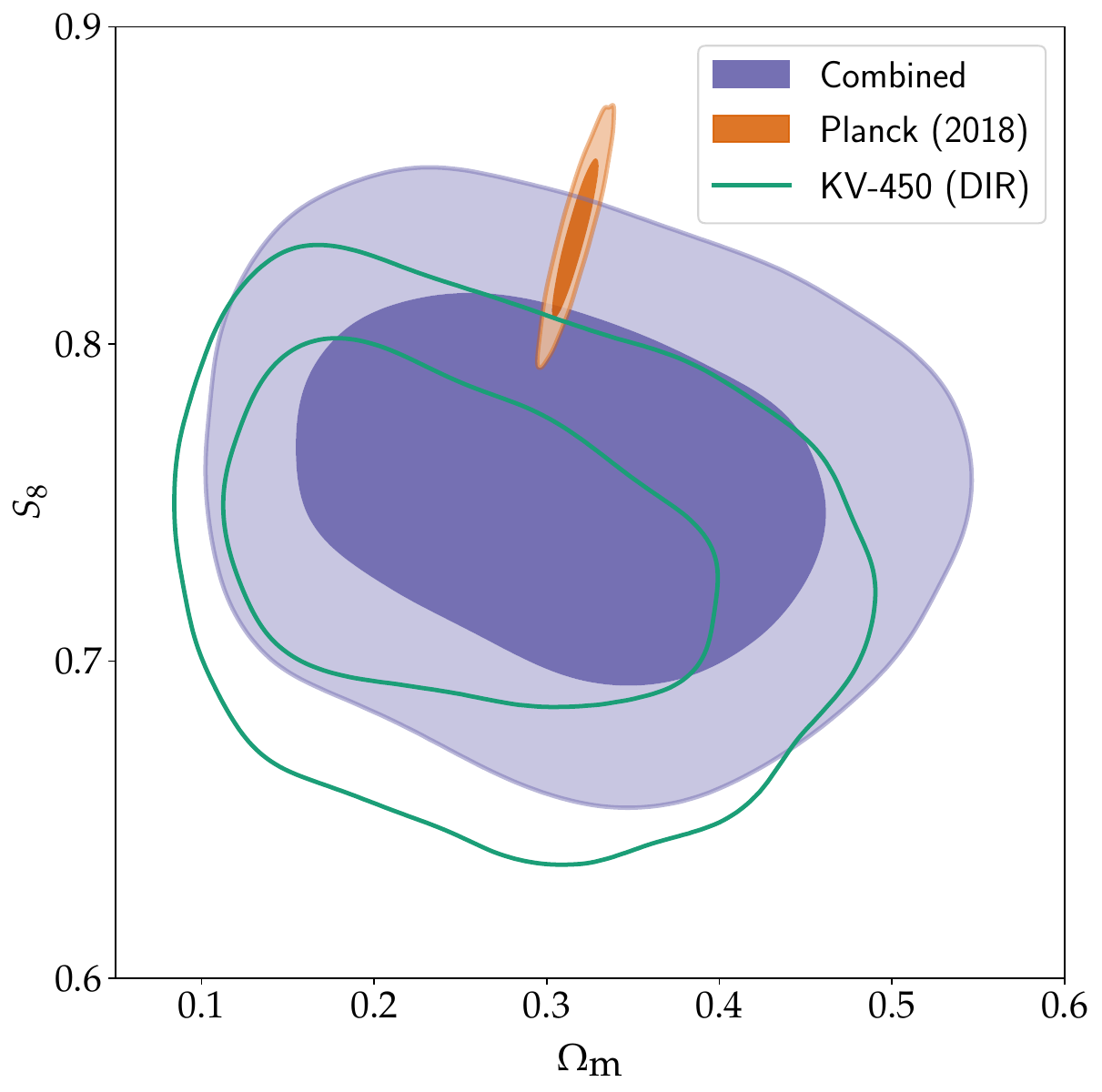}}}
    \hspace{1cm}
    \subfloat[Posterior distribution of $S_{8}$]{{\includegraphics[height=0.30\textheight]{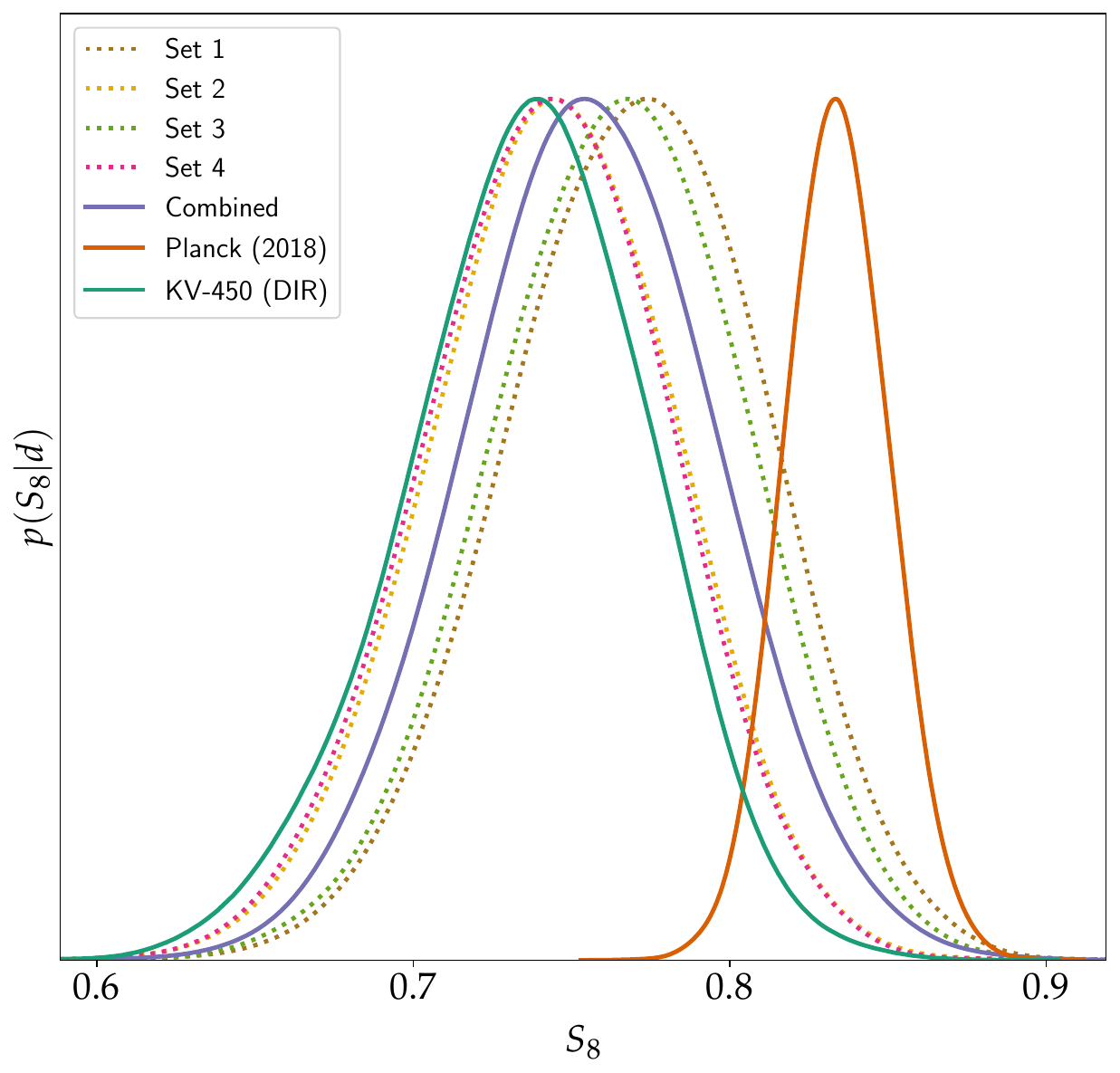}}} 
    \caption{\label{fig:1D_S8}Panel (a): The marginalised posterior distributions (normalised to the same peak height) in the $\Omega_{\tm{m}}-S_{8}$ plane for the KV450 analysis (in purple) and Planck (in orange). The inner and outer contours correspond to the 68\% and 95\% credible intervals respectively. Panel (b): The marginal posterior distribution for the $S_{8}$ parameter in various experiments. The broken curves correspond to the results obtained when different random subsets of $10^5$ galaxies are used to determine $n(z)$.  The the purple curve shows the results when the sets are combined. The distribution of $S_{8}$, as obtained, by the \citet{2020A&A...641A...6P}, is also plotted in orange.}
\end{figure*}

We use standard weak lensing theory (see Appendix B), and include systematics in the same way as \cite{2020A&A...633A..69H} so that direct comparison can be made.  We show in panel (a) of Figure \ref{fig:1D_S8} our $S_{8}-\Omega_{\tm{m}}$ posterior using the combined set of $n(z)$, and compare with that obtained in \cite{2020A&A...641A...6P} and with the DIR KV450 analysis of \cite{2020A&A...633A..69H}.   In Panel (b) of Figure \ref{fig:1D_S8}, we focus only on the 1D marginal posterior distribution of $S_{8}$ and compare the results obtained with $n(z)$ obtained from each galaxy subset (1, 2, 3, 4, and from analysis of the combined set of $4 \times 10^5$ galaxies) with the Planck posterior. As expected, the posterior due to the combined set is roughly centred around the other posterior distributions for KV450, with the results from redshift distributions based on each random set of $10^5$ galaxies showing relatively small sample variance.  The residual sample standard deviation of $S_8$ for the combined sample is approximately 0.006, or a subdominant $0.15\sigma$, 
estimated approximately from the standard deviation of the MAP values divided by $\sqrt{4}$.  This could be reduced with a larger galaxy sample. We find a higher value of $S_{8}=0.756_{-0.039}^{+0.039}$ compared to the original KV450 analysis, reducing the tension with Planck for this dataset.  The error bar is unchanged, as expected, since we use the same summary statistics and cosmology model, and the propagation of the uncertainties in the redshift distributions adds little to the overall error.

\begin{figure*}
\noindent \begin{centering}
\includegraphics[width=0.9\textwidth]{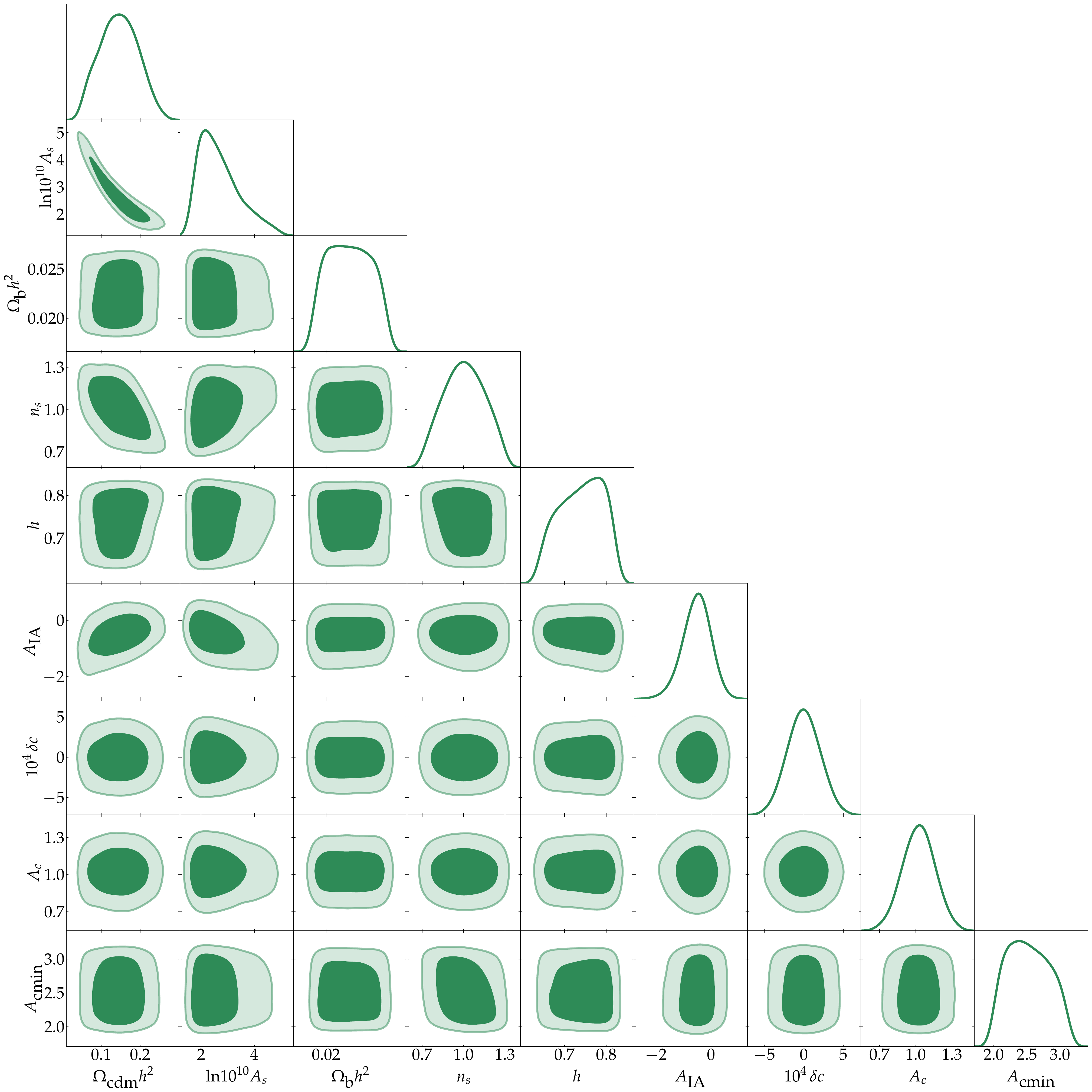}
\par\end{centering}
\caption{\label{fig:triangle_plot} The full marginalised 1D and 2D posterior distribution of both sets of cosmological and nuisance parameters. The inner and outer contours correspond to the $68\%$ and $95\%$ credible intervals. For all parameters, a top-hat prior is assumed except for $\delta c$ and $A_{c}$, for which Gaussian priors are adopted. We refer the reader to Table \ref{tab:results} for further details on the choice of priors and the numerical results with their associated uncertainties.}   
\end{figure*}

In Figure \ref{fig:triangle_plot}, for completeness we show the full 1D and 2D marginal posterior distribution of all the cosmological and nuisance parameters. The inner and outer contours are plotted at 68\% and 95\% credible intervals respectively. For most parameters, a flat uniform prior is assumed except for $\delta c$ and $A_{c}$ for which Gaussian priors are assumed. The mean and $1\sigma$ credible intervals are also reported in Table \ref{tab:results}, but note that the posteriors for most of the parameters apart from $S_8$ are prior-dominated.

While most of the results from weak lensing analyses tend to agree with each other, there remains the tension with the value of $S_{8}$ from Planck, although in this work, this tension is reduced, from 2.3$\sigma$ to 1.9$\sigma$, with an unchanged error bar.

\begin{table*}

\noindent \begin{centering}
\caption{\label{tab:results}Cosmological parameters from KV450 analysis with the new $n(z)$ redshift distributions determined by the Bayesian Hierarchical model.  Note that, with the exception of $S_8$,  many of the formal errors are determined by the priors chosen.}
\renewcommand\arraystretch{1.40}%
\begin{tabular}{lccc}

\textbf{Parameter} & \textbf{Symbol} & Prior& \textbf{Value}\tabularnewline
\hline 
CDM density & $\Omega_{\tm{cdm}}h^{2}$ & $\mc{U}[0.01, 0.99]$ & $0.15_{-0.05}^{+0.05}$\tabularnewline
Scalar spectrum amplitude & $\tm{ln}10^{10}A_{s}$ & $\mc{U}[1.7, 5.0]$ & $2.70_{-0.74}^{+0.79}$\tabularnewline
Baryon density & $\Omega_{\tm{b}}h^{2}$ & $\mc{U}[0.01875, 0.02625]$ & $0.022_{-0.002}^{+0.002}$\tabularnewline
Scalar spectral index & $n_{s}$ & $\mc{U}[0.7, 1.3]$ & $1.01_{-0.15}^{+0.16}$\tabularnewline
Hubble parameter & $h$ &  $\mc{U}[0.64, 0.82]$& $0.74_{-0.06}^{+0.06}$\tabularnewline
\hline 
IA amplitude & $A_{\tm{IA}}$ & $\mc{U}[-6.0, 6.0]$& $-0.53_{-0.49}^{+0.49}$\tabularnewline
Baryon feedback amplitude & $c_{\tm{min}}$ & $\mc{U}[2.00, 3.13]$& $2.5_{-0.3}^{+0.4}$\tabularnewline
Constant $c-$term offset & $10^{4}\delta c$ & $\mc{U}[-6.0, 6.0]$& $-0.03_{-2.0}^{+2.0}$\tabularnewline
2D $c-$term amplitude & $A_{c}$ & $\mc{U}[0.62, 1.40]$& $1.0_{-0.1}^{+0.1}$\tabularnewline
\hline 
Matter density & $\Omega_{\tm{m}}$ & Derived & $0.31_{-0.10}^{+0.10}$\tabularnewline
$S_8=\sigma_{8}\sqrt{\Omega_{\tm{m}}/0.3}$ & $S_{8}$ & Derived & $0.756_{-0.039}^{+0.039}$\tabularnewline
\hline 
\end{tabular}
\par\end{centering}
\end{table*}

%% file: Content/8-Conclusion.tex
\section{Conclusions}
\label{sec:conclusions}

In this paper, we have adapted and developed the Bayesian method of \cite{Leistedt2016} into a practical procedure to infer redshift distributions of galaxies from photometric surveys and a set of templates, by explicitly sampling the redshifts of the galaxies observed in the survey, and applying the method to a real survey in the form of the KV450 dataset. We have also introduced a PCA-based method to reduce the large set of templates to a manageable representative sample, and show that this is effective in providing robust results with around 50 templates.  We have investigated sensitivity to the template choices, and to the subset of $10^5$ source galaxies used to determine the redshift distributions.  Tests on simulated data show that the inferred redshift distributions are more accurate than those from stacked likelihoods.

We have applied this technique in a proof-of-concept study, inferring cosmological parameters from the shear two-point correlation functions from the KV450 data products. We keep the same summary statistics, changing only the redshift distributions, which were originally generated using DIR technique. We marginalise over the $n(z)$ by drawing at random from the samples of $n(z)$ at each likelihood evaluation. This is in contrast to the original KV450 analysis which introduced offsets in the $n(z)$ distributions along the redshift axis as nuisance parameters. Our analysis yields a slightly higher value of $S_{8}=0.756^{+0.039}_{-0.039}$ compared to the original KV450 analysis, which reports a value of $S_{8}=0.737^{+0.040}_{-0.036}$, reducing the tension with Planck from 2.3$\sigma$ to 1.9$\sigma$.

The tension in $S_{8}$ between the weak lensing analyses and the Planck analysis is reduced with the latest KiDS Legacy results \citep{KiDSLegacy}, with the main change coming from the change in the way the redshift distributions are calculated, highlighting the importance of $n(z)$ in cosmological inference. Without spectroscopic redshifts for all sources, all methods suffer from some limitation, such as incomplete spectroscopic coverage, reliance on theoretical models for galaxy spectra, or, in this case, sensitivity to templates, for which a set with broader colour coverage would be necessary for accurate determination for future catalogues. As perhaps emphasized by the KiDS Legacy results, it is helpful to have a range of techniques available in order to assess the robustness of the cosmological results to the approach taken to determine the redshift distributions.  This method gives an alternative approach.

It is worth placing our approach in a broader context. A fully rigorous treatment of cosmic shear would account for the three-dimensional variation of source galaxy number density across the survey volume: the shear field is sampled not uniformly but in proportion to the local galaxy number density, which traces the underlying density field. Consistently accounting for this in both the estimator and the theory prediction would require weighting the lensing kernel to include the effect of the three-dimensional number density field, a level of treatment that is not typically used in current cosmic shear analysis. The standard approach uses a smooth, sky-averaged population-level average redshift distribution. 
Our approach infers the sample redshift distributions from the observed galaxies, and captures the sky-averaged variation due to clustering. However, the main effect of source clustering is in the angular directions \citep{Linke2025}, which is not accounted for in our approach (nor in current analyses), and is an important direction for future work, but note that it is not a major effect, even at the expected precision of Euclid \citep{Linke2025}.

%% file: Content/acknowledgement.tex
\section*{Acknowledgments}

We thank Boris Leistedt and Daniel Mortlock for informative discussion. We also thank Hendrik Hildebrandt and Fabian K\"{o}hlinger for the continuous advice regarding the workings of KiDS data. GTK acknowledges funding through the STFC. AM is supported through the LSST-DA Catalyst Fellowship project; this publication was thus made possible through the support of Grant 62192 from the John Templeton Foundation to LSST-DA. FL acknowledges financial support from the Agence Nationale de la Recherche (ANR) through grant INFOCW, under reference ANR-23-CE46-0006-01. This analysis is based on data products observations made with ESO Telescopes at the La Silla Paranal Observatory under programme IDs 177.A-3016, 177.A-3017, 177.A-3018, 179.A-2004, 298.A-5015, and on data products produced by the KiDS consortium. 

%% file: Content/data_availability.tex
\section*{Data Availability}
The data underlying this article are available from the KiDS website at \href{https://kids.strw.leidenuniv.nl}{https://kids.strw.leidenuniv.nl}. The templates used are from \href{https://www.stsci.edu/~dcoe/BPZ/}{https://www.stsci.edu/~dcoe/BPZ/}.

%% file: Content/9-Appendices.tex
\section*{Appendix A: Generation of simulated data}
\label{app:generation}

For completeness, we include here the details of the generation of the simulated data used for testing. 

The KV450 summary statistics are given a weight given by the shape measurement algorithm {\it lens}fit \citep{lensfit,lensfit2}, so in order to simulate realistically the error distribution and {\it lens}fit weight distribution, a neural network ({\tt{MLPRegressor}} from {\tt{sklearn}}) was trained with one hidden layer of 100 nodes, with input fluxes and output errors or {\it lens}fit weights.  The network was then used to generate simulated flux errors and {\it lens}fit weights for the simulator. Based on the allocated true $z,t,m$, estimated fluxes in the 9 filters are then drawn directly from the model given by Equations \ref{eq:fluxerror} and \ref{eq:meanflux}.

Simulated sources were allocated to tomographic bins using the KV450 BPZ pipeline, converting to magnitudes and magnitude errors according to  $m_b=-2.5\log_{10}(\hat{F}_b)$ and error $m_{e,b} = 1.086 \times \sigma_b/\hat{F}_b$.
Sources have filters classed as `unobserved' if they fall under one of two conditions, either $\sigma_b > \hat{F}_b$ or $m > b_{\rm lim} $ where $b_{\rm lim}$ is the magnitude limit for a certain panel of sky. These limits were randomly allocated to sources. 
Although technically these sources are `undetected', they are classed as unobserved.  This methodology is replicated from the original KV450 BPZ results which found that using `unobserved' over `undetected' led to fewer extreme outliers, and is kept to keep consistency with the KiDS analysis, so we can use the same summary statistics. The classification of `unobserved' means setting $m=99$ and BPZ then treats the filter to have maximum error and no information can be gathered from it. BPZ also takes a $M_0$ input, which in KV450 data is labelled `MAG\_AUTO' and consists of a nonlinear combination of r and i magnitudes and their errors. The $M_0$ value is used in the BPZ prior. $M_0$ values were also simulated using a trained neural network where $m_{\rm r,i}$ are the inputs and $M_0$ the output.  Note that this only affects the assignment to tomographic bins; it is not the correct procedure for inference and is not used in the BHM.

The BPZ point estimate $\hat{z}_{\tm{BPZ}}$ was then used to allocate sources to 5 tomographic bins in the range $z=0.1$ to 1.2, with internal boundaries at (0.3,0.5,0.7,0.9).  Sources with estimated redshifts outside this range were discarded. The simulated surveys were analysed using the same PCA template set as the main survey.

\section*{Appendix B: weak lensing data model}
\label{app:theory}

In this appendix, we review the weak lensing theory used, and define the summary statistics used in the analysis. Since it is relatively standard, we place it here.
To effect a direct comparison, we replicate the KV450 analysis, but with a new set of redshift distributions, which we marginalise over. 
All other data products, such as the data vector and the covariance matrix are the same as those by \citet{2020A&A...633A..69H} to which we refer the reader for further details.

\subsection*{Cosmic shear summary statistics}
\label{sec:cosmo_data}

Here we define the summary statistics used, which are the same as in the original KV450 analysis, to allow a direct comparison of results.  KV450 employs the TREECORR code to estimate the shear correlation function between two tomographic bins $\alpha$ and $\beta$, that is, 
\begin{equation}
    \hat{\xi}_{\pm}^{\alpha\beta}(\theta)=\dfrac{\sum_{pq}w_{p}w_{q}\left[\epsilon_{\tm{t}}^{\alpha}(\bs{x}_{p})\epsilon_{\tm{t}}^{\beta}(\bs{x}_{q})\pm\epsilon_{\times}^{i}(\bs{x}_{p})\epsilon_{\times}^{j}(\bs{x}_{q})\right]}{\sum_{pq}w_{p}w_{q}}
\end{equation}
\noindent where $w$ is the \textit{lensfit} weight, $\epsilon_{\tm{t}}$ and $\epsilon_{\times}$ correspond to the tangential and cross ellipticities of a galaxy with reference to the vector $\bs{x}_{p} - \bs{x}_{q}$ between a pair of galaxies $p$ and $q$. Nine logarithmically spaced bins are used within the interval $\theta\in [0'.5,\,300']$. We follow the same procedure as \citet{2020A&A...633A..69H} and use the first seven bins for $\xi_{+}$ and last six bins for $\xi_{-}$. Since we have five tomographic redshift distributions, we have five auto and ten cross pairs to tomographic redshift bins, the total number of data points is $(10+5)\times(7+6)=195$,
 
and use the data covariance matrix of \citet{2020A&A...633A..69H}.

In \S\ref{sec:theory}, we cover the forward theoretical model and throughout this work, for consistency with the original KV450 analysis we assume a Gaussian likelihood for the data with a fixed covariance matrix.

\subsection*{Theory}
\label{sec:theory}
To model the data, the two-point correlation functions between bins $\alpha$ and $\beta$ are used as the forward cosmological model. It is strictly a linear combination of the E- and B-mode power spectra, that is, 
\begin{equation}
\label{eq:shear_corr}
\xi_{\pm}^{\alpha\beta}(\theta)=\int_{0}^{\infty} \dfrac{\ell}{2\pi}J_{0/4}(\ell\theta)[C_{\ell,\alpha\beta}^{\tm{EE}}\pm C_{\ell,\alpha\beta}^{\tm{BB}}]
\end{equation}
\noindent where $J_{0/4}$ are  Bessel functions of the first kind.  In the absence of systematics, the B-mode power spectra are assumed to be zero and the main contribution is due to the convergence power spectra, $C_{\ell,\alpha\beta}^{\tm{EE}}=C_{\ell,\alpha\beta}^{\kappa\kappa}$. Under the Limber approximation \citep{2008PhRvD..78l3506L} and assuming a flat universe, the EE power spectrum is given by
\begin{equation}
C_{\ell,\alpha\beta}^{\tm{EE}}=\int_{0}^{\chi_{\tm{H}}}\tm{d}\chi\;\dfrac{w_{\alpha}(\chi)\,w_{\beta}(\chi)}{\chi^{2}}\,P_{\delta}(k;\chi)
\end{equation}
\noindent where $P_{\delta}(k;\chi)$ is the non-linear matter power spectrum and $\chi$ the comoving distance. The weight function $w_\alpha(\chi)$ for tomographic bin $\alpha$ is
\begin{equation}
\label{eq:lensing_efficiency}
w_{\alpha}(\chi)=A\chi(1+z)\int_{\chi}^{\chi_{\tm{H}}}\tm{d}\chi'\,n_{\alpha}(\chi)\left(\dfrac{\chi'-\chi}{\chi'}\right)
\end{equation}
\noindent and $A=\frac{3H_{0}^{2}\Omega_{\tm{m}}}{2c^{2}}$. The weight functions depend on the redshift distribution, $n_{\alpha}(z)\,dz=n_{\alpha}(\chi)\,d\chi$, which are here normalised such that
\begin{equation*}
    \int n(z)\,dz = 1.
\end{equation*}
\noindent In this analysis, we sample five cosmological parameters
\begin{equation*}
    \Phi = \left[\Omega_{\tm{cdm}}h^{2},\,\tm{ln}10^{10}A_{s},\,\Omega_{\tm{b}}h^{2},\,n_{s},\,h\right],
\end{equation*}
where $\Omega_{\tm{cdm}}$ is the CDM density parameter,   $A_{s}$ is the power spectrum amplitude, $\Omega_{\rm b}$ is the baryon density (so $\Omega_{\tm{m}}=\Omega_{\tm{cdm}}+\Omega_{\rm b})$, $n_{s}$ is the scalar spectral index, and $h$ is the Hubble constant in units of $100$\,km\,s$^{-1}$\,Mpc$^{-1}$. 
\noindent We assume two massless neutrinos and one massive neutrino are assumed, with the latter fixed at $m_{\nu}=0.06\,\tm{eV}$. We also record two derived quantities, namely $\Omega_{\tm{m}}$ and $S_{8}\equiv \sigma_{8}\sqrt{\Omega_{\tm{m}}/0.3}$. Sampling is done in \texttt{MontePython} \citep{2019PDU....24..260B}, which also wraps \texttt{CLASS} \citep{2011JCAP...07..034B}. Apart from the cosmological parameters, we also have to account for nuisance parameters, which we discuss briefly in the next section.

\subsection*{Systematics}
\label{sec:systematics}
The recovered $C_{\ell}$ is a biased tracer of the convergence power spectrum. The model which is commonly used in recent weak lensing analyses incorporates additional terms due to intrinsic alignments, II and GI, that is,
\begin{equation}
\label{eq:total_cl}
    C_{\ell,\alpha\beta}^{\tm{tot}}=C_{\ell,\alpha\beta}^{\kappa\kappa} + A_{\tm{IA}}^{2}C_{\ell,\alpha\beta}^{\tm{II}}+A_{\tm{IA}}C_{\ell,\alpha\beta}^{\tm{GI}}.
\end{equation}
\noindent The II term which arises due to correlation of ellipticities in the local environment, contributes positively to the total lensing signal, while the GI term arises due to correlation between intrinsic ellipticities of foreground and gravitational shear of background galaxies \citep{2004PhRvD..70f3526H}. The II power spectrum  is modelled as
\begin{equation}
C_{\ell,\alpha\beta}^{\tm{II}}=\int_{0}^{\chi_{\tm{H}}}\tm{d}\chi\;\dfrac{n_{\alpha}(\chi)\,n_{\beta}(\chi)}{\chi^{2}}\,P_{\delta}(k;\chi)\:F^{2}(\chi)
\end{equation}
\noindent where $F(\chi)=C_{1}\rho_{\tm{crit}}\Omega_{\tm{m}}/D_{+}(\chi)$ \citep{2007NJPh....9..444B, 2011A&A...527A..26J}. $C_{1}$ is a constant given by $5\times10^{-14}\;h^{-2}\tm{M}_{\odot}^{-1}\tm{Mpc}^{3}$, $D_{+}(\chi)$ is the linear growth factor normalised to unity today and $\rho_{\tm{crit}}$ is the critical density of the Universe today. The GI power spectrum is modelled as
\begin{equation}
C_{\ell,\alpha\beta}^{\tm{GI}}=\int_{0}^{\chi_{\tm{H}}}\tm{d}\chi\,\dfrac{w_{\alpha}(\chi)n_{\beta}(\chi)+w_{\beta}(\chi)n_{\alpha}(\chi)}{\chi^{2}}\,P_{\delta}(k;\chi)\,F(k;\chi).
\end{equation}
\noindent Note that $A_{\tm{IA}}$ in Equation \ref{eq:total_cl} is a free amplitude parameter, which is marginalised over in the sampling procedure. Taking into account the two effects due to intrinsic alignment, the shear correlation function can be written as 
\begin{equation}
    \xi^{\tm{tot}}_{\pm}=\xi^{\tm{EE}}_{\pm} + A_{\tm{IA}}^{2}\xi^{\tm{II}}_{\pm} + A_{\tm{IA}} \xi^{\tm{GI}}_{\pm}.
\end{equation}
Another source of theoretical systematics is baryon feedback. It causes a reduction in the matter power spectrum at small scales. In the analysis of KiDS-450 data \citep{2017MNRAS.471.4412K}, baryon feedback was modelled using a fitting formulae as derived by \citet{2015MNRAS.450.1212H}. It is worth highlighting that these processes are not very well understood and hydrodynamical simulations were not strictly carried out with a view to matching them to weak lensing observations, but rather to understand the processes behind galaxy formation and feedback effects.

In this work, following the same approach as in \citet{2020A&A...633A..69H} for a comparative study, we use HMCode \citep{2015MNRAS.454.1958M} where the halo bloating parameter, $\eta=0.98-0.12c_{\tm{min}}$ changes the halo density profile. $c_{\tm{min}}\in[2.0,\,3.13]$ is a free parameter which is marginalised over in the likelihood analysis. A value of $c_{\tm{min}}=2.0$ corresponds to an extreme feedback model (AGN) while $c_{\tm{min}}=3.13$ corresponds to dark matter only.

In addition to the above nuisance parameters, two additional parameters, $\delta c$ and $A_{c}$, were introduced to account for the additive shear bias and position dependent additive bias pattern respectively. In short, the set of nuisance parameters, which we denote as $\Psi$ is (see \cite{2020A&A...633A..69H} for full details):
\begin{equation*}
    \Psi = \left[A_{\tm{IA}}, c_{\tm{min}},\,\delta c, A_{c}\right].
\end{equation*}
The new $n(z)$ method developed in this work is then coupled with the KV450 likelihood and the cosmological and nuisance parameters are inferred.

\section*{Appendix C: KV450 and template colours}

\begin{figure*}
\noindent \begin{centering}
\includegraphics[width=.95\textwidth]{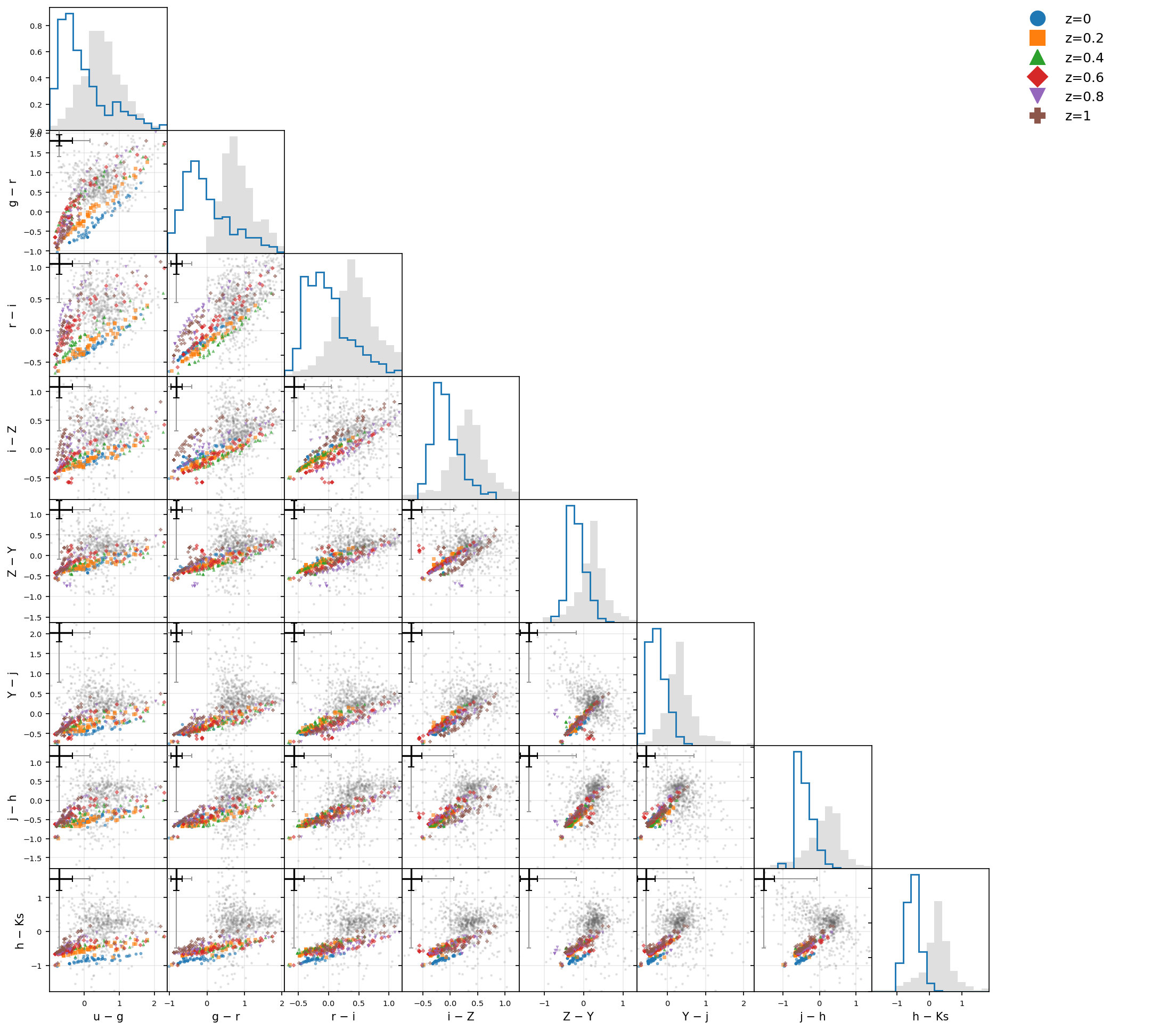}
\par\end{centering}
\caption{\label{fig:colours} The distribution of colours of a random subset of KV450 galaxies (in grey) and the PCA templates.  Median and maximum errors are shown by the error bars.}   
\end{figure*}

In Fig. \ref{fig:colours} we show the distribution of KV450 colours and the PCA template colours.  The templates are shown at different redshifts roughly spanning the range of KV450 redshifts.  The diagonals show the prior distributions of template colours at the indicated redshifts.  The median and maximum KV450 galaxy errors are shown in each subfigure. 1\% of colour outlier galaxies are outside the limits for display purposes.

